\documentclass[]{article}

\usepackage{latexsym}
\usepackage{epsfig}

\begin{document}

\baselineskip=5mm \quad

\vspace{8pt} \centerline{\Large \bf Volume Effect of Bound States
in Quark Gluon Plasma}

\vspace{32pt} \centerline{ \large Hong Miao$^1$, Chongshou
Gao$^{2,3}$, Pengfei Zhuang$^1$}

\vspace{16pt} \centerline{ $^1$Physics Department, Tsinghua
University, Beijing 100084, China}

\vspace{4pt} \centerline{ $^2$School of Physics, Peking
University, Beijing 100871, China}

\vspace{4pt} \centerline{ $^3$Institute of Theoretical Physics,
Academia Sinica, Beijing 100080, China}

\normalsize

\vspace{16pt}
\begin{abstract}

Bound states, such as $qq$ and $q\bar{q}$, may exist in the Quark
Gluon Plasma. As the system is at high density, the volume of the
bound states may evoke a reduction to the phase space. We
introduce an extended bag model to investigate qualitatively the
volume effect on the properties of the system. We find a limit
temperature where the bound states start to be completely melted.

\vspace*{8pt}

{\bf PACS number(s)\/}: 12.38.Mh 25.75.-q


{\bf Key words\/}: \textit{QGP, Equation of State, Diquark, Bag
Model, Volume Effect}

\end{abstract}

\vspace{22pt}

\leftline{\large \bf 1. Introduction}

\vspace{8pt}

It is discussed recently that the Quark Gluon Plasma (QGP) may
contain bound states, especially binary bound states such as $qq$,
$q\bar{q}$, $gq$ and $gg$ at temperatures above the critical value
of deconfinement phase transition $T_c$ \cite{sign:sign
diquark_baryon}\cite{sign:sign diquark_density}\cite{sign:sign
bound_state}. Inserting these new degrees of freedom into the QGP
will change its properties such as pressure and energy density.
From lattice calculations \cite{sign:sign QuasiParticle1},  the
masses of quarks and gluons are very heavy, such as
$m_{q,g}/T\sim3$ at $T\approx1.5-2 T_c$. If the bound states with
finite volumes are introduced to QGP as well as the point like
quarks and gluons with current masses, the high pressure at high
$T$ will be easy to be reached.

The influence of particle volume is widely discussed in the Van
der Waals Excluded Volume Model (VdW) \cite{sign:sign
WaalsSPS}\cite{sign:sign WaalsHG} to describe hadronic gas at high
density. In QGP, the number density is very high so that the
ball-like bound states would be quite close to each other as
discussed by J. W. Clark, J. Cleymans and J. Rafelski
\cite{sign:sign_vol_lit}. The volume effect of bound states may
change the properties of such strongly coupled system remarkably.

In this paper, detailed calculations will be presented to describe
the influences of binary bound states in QGP. This may be helpful
for us to understand the Equation of State (EOS) of QGP, which is
quite important for the study of relativistic heavy ion
collisions\cite{sign:sign EOS}. Many-body bound states such as
baryons may also exist\cite{sign:sign byinQGP1}, but they are
neglected here for their large masses.

\vspace{10pt}

\leftline{\large \bf 2. Influence of the Bound State Volume}

\vspace{8pt}

Generally, statistical values such as $\varepsilon$ and $P$ could
be described as corresponding integrals,
 \begin{equation}
 F_i = \frac{1}{(2\pi)^3V}\int{Y_{Fi}(p)f_{i}(p)d^3pd^3r},
 \end{equation}
where $f_{i}(p)$ is the momentum distribution of number density
for $i$ particles. If all the components are point particles, the
integral of coordinates equals to the system volume. When
particles with excluded volume exist in the system, some of the
coordinate space will be occupied,
 \[
 \int{d^3r} < V,
 \]

Regarding those particles as rigid balls, the integral can
described as,
 \begin{equation}
 \int{d^3r}=(1-y_{op})V,
 \end{equation}
where $y_{op}$ is the ratio of the volume occupation of the
components. Then, for the bound states with volumes, one has
 \begin{eqnarray}
 \nonumber
 F_i & = & (1-y_{op})F_{i0}\\
                & = & (1-y_{op})\frac{1}{(2\pi)^3}\int{Y_{Fi}(p)f_{i}(p)d^3p},
 \label{eqs:bound}
 \end{eqnarray}

This situation is very similar to the equation of Van der Waals in
terms of the influence of particle volumes, $(V-Nb)$, in
 \begin{equation}
 \left[P+\left(\frac{N}{V}\right)^2a\right](V-Nb)=NkT.
 \end{equation}
In relativistic conditions, $y_{op}$ is defined as,
 \begin{equation}
 y_{op} = (1-y_{op})y_{op0}.
 \end{equation}
Then,
 \begin{equation}
 y_{op}=\frac{y_{op0}}{1+y_{op0}},
 \label{eqs:yop}
 \end{equation}
where\cite{sign:sign Strange},
 \begin{equation}
 y_{op0}=\sum_{i}{\frac{1}{(2\pi)^3}\int{\frac{m_i}{E_i}V_{i}f_{i}(p)d^3p}},
 \label{eqs:yop0}
 \end{equation}
Here, $V_{i}$ is the rest volume of particle $i$ depending on
temperature or energy density. $\frac{m_i}{E_i} = \gamma_i^{-1}$
is the factor of Lorenz-contract.

For point particles such as quarks and gluons, the "rigid balls"
are transparent in fact. That means quarks and gluons can pass
through the balls, but not completely freely as in vacuum. This
situation is equivalent to a reduction of phase space, especially
in low momentum regions.

As an assumption, a parameter $C_\alpha\leq 1$ is used to describe
the transparency for all the balls. $C_\alpha$ is supposed to be a
function of $T$ or other variables, but the details are not clear.
Approximately it would remain constant at given temperatures and
is regarded as an adjustable parameter in our calculations to
provide a preliminary demonstration. As a comment, the scattering
of quarks and gluons with the bound state, which would destroy
some of the bound states, will occur in the same time. This effect
is part of the chemical equilibrium in QGP.

Then, for point particles,
 \begin{equation}
 F_i = \left[\;(1-y_{op})+C_\alpha y_{op}\;\right]F_{i0}.
 \end{equation}

Specially, $P_{20}$ and $\varepsilon_{20}$ are used to describe
the pressure and energy density of binary bound states and
$P_{10}$, $\varepsilon_{10}$ are used to describe those of quarks
and gluons. Then for the whole system,
 \begin{eqnarray}
 \nonumber
 P              & = & (1-y_{op})P_{20} + [\;(1-y_{op})+C_{\alpha}y_{op}]P_{10},    \\
 \varepsilon    & = & (1-y_{op})\varepsilon_{20} + [\;(1-y_{op})+C_{\alpha}y_{op}]\varepsilon_{10}.
 \label{eqs:point}
 \end{eqnarray}

In order to study the influence on Quark Gluon Plasma when bound
states with volumes are considered, the expressions of masses and
volumes of those binary bound states and their dependence on $T$
or $\varepsilon$ are needed. Generally, bound state masses would
increase with temperature or energy density. This property makes
them fewer and fewer, less stable and less important in QGP at
high temperatures \cite{sign:sign bound_state}. Volumes are also
supposed to increase at the same time. A phenomenological model is
required to construct masses and volumes in such styles. One of
the suitable models to build volume is the Bag
Model\cite{sign:sign MIT_Bage}\cite{sign:sign_vol_lit}. It will be
extended approximately in the condition of finite temperature.

\vspace{10pt}

\leftline{\large \bf 3. Extended  Bag Model}

\vspace{8pt}

MIT Bag Model\cite{sign:sign MIT_Bage} describes the quark
confinements in hadrons by additional boundary conditions. The
mass of a hadron is expressed as \cite{sign:sign Bag_mass},
 \begin{equation}
 M(R)=E_V + E_0 + E_Q + E_M + E_E,
 \label{eqs:bag_mass}
 \end{equation}
where $E_V = BV_B$ is the volume energy, $E_0=-Z_0/R$ is the
zero-point energy with constant $Z_0$ and $E_Q$ is the
contribution of the confined quarks in the spherical region. $E_Q$
will decrease to the sum of quark masses when the radius of the
bag expands. $E_M$ and $E_E$ are the color magnetic exchange
energy and the color electric interaction energy. By the quadratic
boundary condition of pressure equilibrium, which is equivalence
to minimizing $M(R)$ with respect to $R$. Then, the bag radius
$R_0$ can be determined. Bag parameter was fit to be $B^{1/4}=145
MeV$\cite{sign:sign Bag_mass}.

Bag model is originally used to describe the free hadrons in
vacuum. When dense medium at high temperature is considered, the
model need to be extended. Some discussions based on GCM or NJL
model in conditions of finite density \cite{sign:sign
Bag_Liu}\cite{sign:sign Bag_Burgio}\cite{sign:sign Bag_Aquirre}
noted that $B$ would decrease with temperature or medium density.

On the other hand, when discussing the phase transition between
hadron matter and QGP, $B$ is also introduced to balance the
pressure difference between different phases. The value is fit to
be $B^{1/4}=230MeV$\cite{sign:sign hydro1} at $Tc=164MeV$, and
there is an approximate relation of $B \propto Tc^4$.

These two directions can hardly meet together. It is also noted
that $B$ is originally defined to be both energy density
difference and pressure difference between the opposite sides of
the bag boundary. When medium exists, they will not be the same
any more. The linear boundary conditions and the quadratic
boundary conditions\cite{sign:sign MIT_Bage} will also change. It
means the model requires a small extension.

In the extension, $B\equiv B_0$ is only regarded as the
contributions of the sea quarks and gluons in basic states, which
on whole contain zero quantum numbers. $B_0$ is same to that in
the original MIT Bag Model in vacuum. Keeping the pressure
equilibrium or distinguishing different vacuum types between both
sides of bag boundary by $B$ of the same value may not be needed,
as the color confinement may not exist in QGP. After all, this
extension is just used to carry out a mechanism to discuss volumes
and masses. The minimizing condition now is used on the whole
energy of the system including medium instead,
 \begin{eqnarray*}
 E(R)   &=& E_{Bag} + E_{med} \\
        &=& B_0V_B + E_0 + E_Q + E_M + E_E + \varepsilon_{medout}(V_0-V_B) + \varepsilon_{medin}V_B,\\
        &=& [B_0- (\varepsilon_{medout} - \varepsilon_{medin})]V_B + E_0 + E_Q + E_M + E_E + \varepsilon_{medout}V_0,
 \end{eqnarray*}
where $V_0$ is the system volume and ($\varepsilon_{medout}V_0$)
is constant to $R$. At the same time, the effective bag parameter
$B_{eff}=B_0-(\varepsilon_{medout} - \varepsilon_{medin})$
decreases, which is consistent with the opinions of
\cite{sign:sign Bag_Liu}\cite{sign:sign Bag_Burgio}\cite{sign:sign
Bag_Aquirre}. While, the bag mass is still defined as Eq
(\ref{eqs:bag_mass}) with $E_V = B_0V_B$. Direct dependence on
chemical potentials is neglected.

However, it is noted that $\varepsilon_{medout}$ and
$\varepsilon_{medin}$ are not the summation of the energy per unit
volume for the particles located inside or outside the bag
boundary. That is because the volume effect can only restrict the
coordinate space of the particle barycenters, instead of the
particle wave functions or mass distributions. So, the particles
located in one side of the boundary could contribute the energy
density appropriately to the other side. Qualitatively,
 \begin{eqnarray*}
 \varepsilon_{medin}  &=&
    \frac{1}{V_{in}} \int_{in}
    {[\sum_{in}^{barycenter}{\varepsilon_i^{\textrm{I}}\Psi_i^{\textrm{I}2}(\mathbf{x})}
    +\sum_{out}^{barycenter}{\varepsilon_i^{\textrm{II}}\Psi_i^{\textrm{II}2}(\mathbf{x})}]}d^3x, \\
 \varepsilon_{medout} &=&
    \frac{1}{V_{out}} \int_{out}
    {[\sum_{in}^{barycenter}{\varepsilon_i^{\textrm{I}}\Psi_i^{\textrm{I}2}(\mathbf{x})}
    +\sum_{out}^{barycenter}{\varepsilon_i^{\textrm{II}}\Psi_i^{\textrm{II}2}(\mathbf{x})}]}d^3x,
 \end{eqnarray*}
where, $\varepsilon^{\textrm{I}}$ and $\varepsilon^{\textrm{II}}$
are the energy per unit volume for the particles located inside
and outside the bag boundary, $\Psi$ is the effective wave
function. Considering that the energy density difference between
boundaries would be much smaller than expected by the
superimposition of particle wave functions, a rest factor $c_{ij}$
could be introduced as a kind of average for a specific bag $j$,
 \[
 (\varepsilon_{medout} - \varepsilon_{medin})_j
    \approx \sum_{i}{(\varepsilon_i^{\textrm{II}}c_{ij}^{\textrm{II}}-\varepsilon_i^{\textrm{I}}c_{ij}^{\textrm{I}})}
    \approx \sum_{i}{(\Delta\varepsilon_i c_{ij})},
 \]

Summation to $i$ and neglecting the difference of $j$, one has
approximately,
 \begin{equation}
 (\varepsilon_{medout} - \varepsilon_{medin})_j
    \approx \Delta\varepsilon_{med} c_{j}
    \approx c\Delta\varepsilon_{med}
    \approx f(\Delta\varepsilon_{med}),
 \end{equation}
where,
 \begin{equation}
 \Delta\varepsilon_{med} = \varepsilon^{\textrm{II}} - \varepsilon^{\textrm{I}} = (1-y_{op})[\varepsilon_{20} + (1-C_{\alpha})\varepsilon_{10}],
 \label{eq:de}
 \end{equation}

$f(\Delta\varepsilon_{med})$ should increase slowly with
$\Delta\varepsilon_{med}$ or $T$, but the detailed expression may
be less important for the final qualitative results like the Ising
model. To avoid the divergence of numerical calculations,
 \[
 c\Delta\varepsilon_{med}
    =       B_0\left(\frac{c\Delta\varepsilon_{med}}{B_0}\right)
    \approx B_0( 1- e^{-\frac{c\Delta\varepsilon_{med}}{B_0}})
    =       B_0 (1-e^{-C_e \Delta\varepsilon_{med}}),
 \]
After defining,
 \begin{equation}
 f(\varepsilon) = B_0 (1-e^{-C_e \varepsilon}),
 \label{eq:fe}
 \end{equation}
the bag masses and volumes could be determined by minimizing
 \begin{equation}
 E(R) = ( B_0-f(\Delta\varepsilon_{med}) )V_B + E_0 + E_Q + E_M + E_E.
 \label{eqs:bagExt}
 \end{equation}

\begin{figure}[htb]
 \centering
  \hspace{0.025\textwidth}%
  \begin{minipage}[t]{0.45\textwidth}
    \centering
    \includegraphics[width=\textwidth]{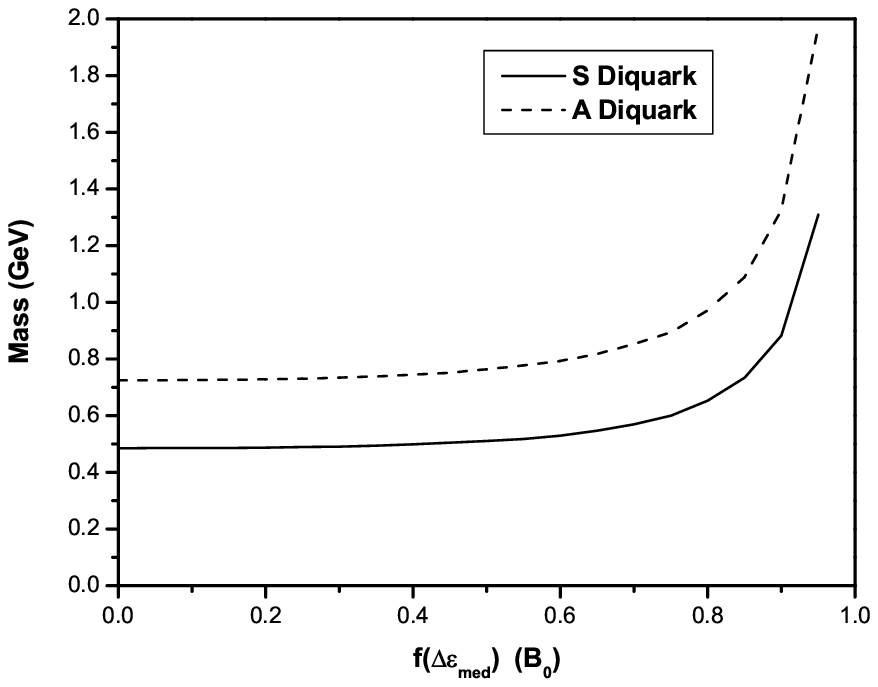}
    \caption{\quad Diquark masses with two light quarks via
    $f(\Delta\varepsilon_{med})$, $B_0$ in unit. The upper curve is
    the mass of axial-vector diquark. The lower curve is
    the mass of scalar diquark. Where, $B_0^{1/4} = 145 MeV$, $Z_0 = 1.84$ and $\alpha_c = 0.55$ as unadjustable constants from \cite{sign:sign Bag_mass}.}
    \label{fig:mdt}
  \end{minipage}%
  \hspace{0.05\textwidth}%
  \begin{minipage}[t]{0.45\textwidth}
  \centering
    \includegraphics[width=\textwidth]{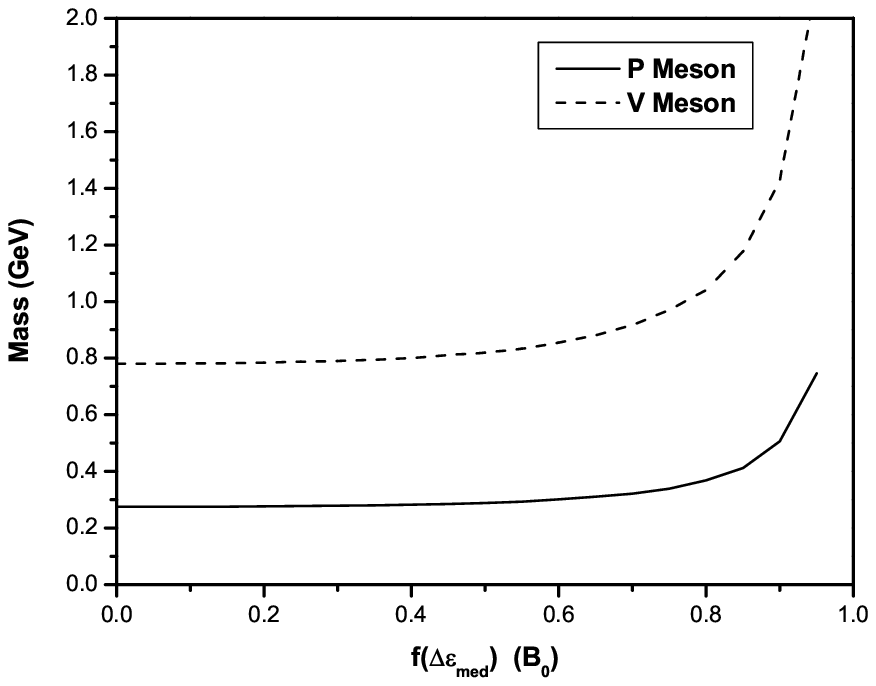}
    \caption{\quad Masses of color singlets ("mesons") with a light quark and a light anti-quark  via
    $f(\Delta\varepsilon_{med})$, $B_0$ in unit. The upper curve is
    the mass of vector bound state. The lower curve is
    the mass of pseudo-scalar bound state. Where, $B_0^{1/4} = 145 MeV$, $Z_0 = 1.84$ and $\alpha_c = 0.55$.}
    \label{fig:mmt}
 \end{minipage}%
  \hspace{0.025\textwidth}
\end{figure}

Masses of some diquarks ($qq$)\cite{sign:sign diquark_form} and
$q\bar{q}$ singlets are shown in Fig (\ref{fig:mdt}) and
(\ref{fig:mmt}). They increase slowly with
$f(\Delta\varepsilon_{med})$, until it is close to $B_0$, which
makes the bag masses expand to infinite. The bag volumes also
increase as the same styles. It is noted that in QGP with bound
states, where the color and chiral symmetry is completely or
partly restored, the pseudo-scalar $q\bar{q}$ singlets may not be
Goldstone Bosons. They are considered much heavier than the pions
in vacuum.

\begin{figure}[htb]
 \centering
  \hspace{0.025\textwidth}%
    \begin{minipage}[t]{0.45\textwidth}
        \centering
        \includegraphics[width=0.91\textwidth]{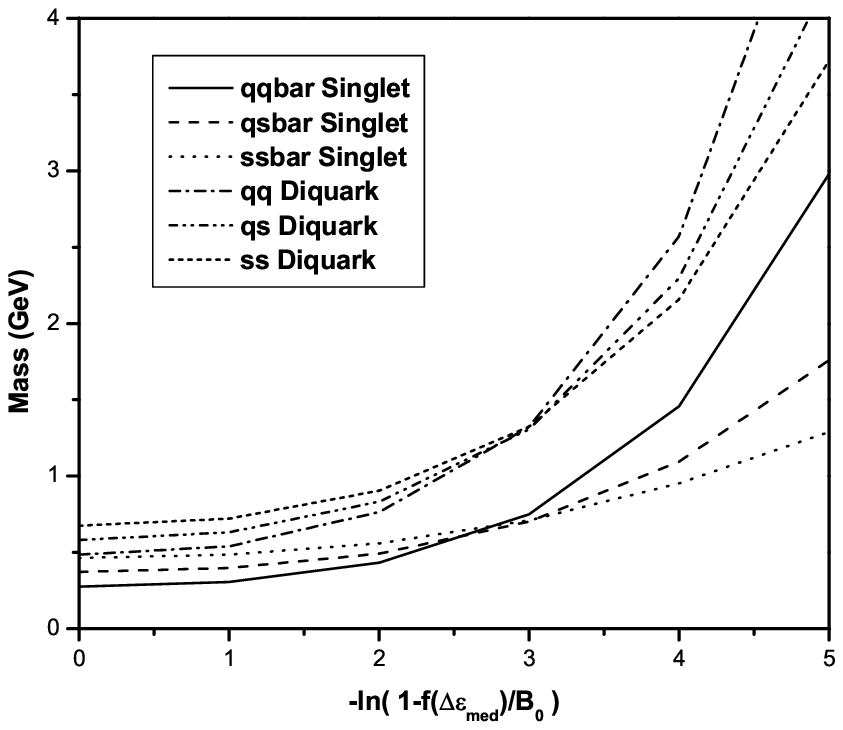}
        \caption{\quad Masses of pseudo-scalar singlets ("mesons") and scalar diquarks with different components via
            $f(\Delta\varepsilon_{med})$. When $f(\Delta\varepsilon_{med})$ is high enough,
            bound states with heavier quarks become smaller than those with light quarks.
        }
        \label{fig:HeavyStable}
    \end{minipage}%
  \hspace{0.05\textwidth}%
  \begin{minipage}[t]{0.45\textwidth}
    \centering
    \includegraphics[width=\textwidth]{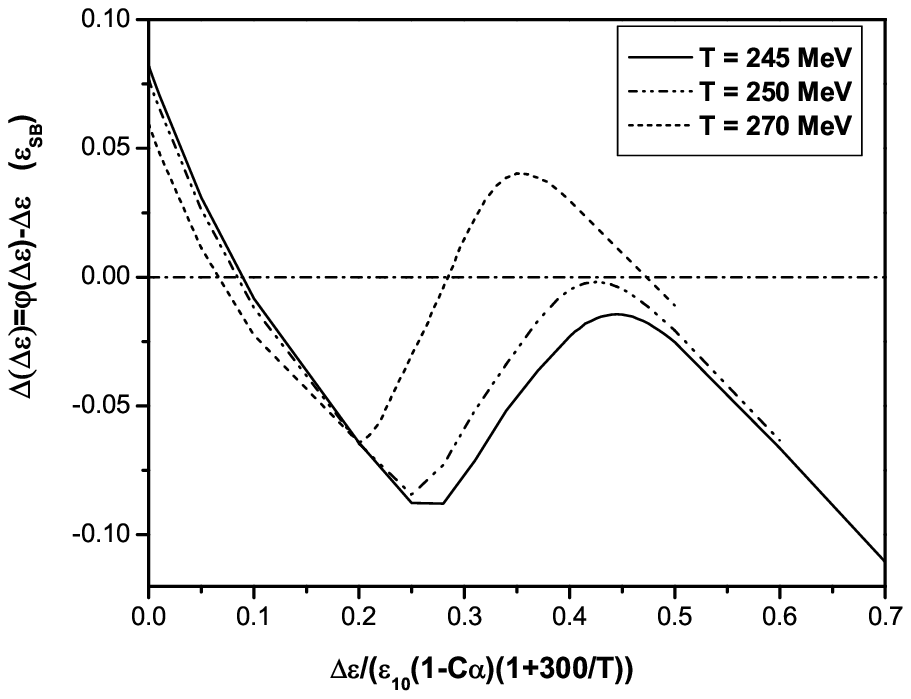}
        \caption{\quad Demonstration of the solutions of the self-consistent equation {(\ref{eqs:self})},
            where $C_\alpha =0.8$ and $C_e =5\times10^{-10}MeV^{-4}$.
            Factor $\varepsilon_{10}(1-C_\alpha)(1+300/T)$ is just used for the numerical
            calculations, where $\varepsilon_{10}$ is the original energy
            density of quarks and gluons without bound
            states, see in Eq (\ref{eqs:point}).
        }
        \label{fig:RootFind}
 \end{minipage}%
 \hspace{0.025\textwidth}
\end{figure}

Although masses of bound states with light quarks are smaller than
those with heavier ones in vacuum, they will be larger when
$f(\Delta\varepsilon_{med})$ grows high enough close to $B_0$.
Demonstrations of $S=0$ bound states are shown in Fig
(\ref{fig:HeavyStable}), $S=1$ bound states act as the same styles
too. This implies that $ss$ and $s\bar{s}$ will be more important
and more stable in conditions of high temperature in QGP relative
to those bound states with light quarks. It is also consistent
with the opinions that heavy quark bound states, such as $J/\Psi$,
will contain larger effective potentials and gain more stabilities
in QGP, although charm quark can not be used in this model so far.

\vspace{10pt}

\leftline{\large \bf 4. Statistical Properties of QGP with Bound
States}

\vspace{8pt}

There are 394 binary bound states  all together for $N_f=2$ system
and 588 for $N_f=3$ system. In our calculations, it is supposed
there is no color SU(3) symmetry breaking in Quark Gluon Plasma.
That means some antisymmetric states with two identical particles
such as $J=1$ $gg$ states and some of pseudo-scalar diquarks do
not take into account.

The system is regarded as an "infinitive" system approximately,
with low chemical potentials. But leptons, which will escape from
the finite region created by the relativistic heavy ion
collisions, are neglected.

Two light quarks and a heavy strange quark is used in the
discussions,
 \[
 m_{g} = 0 MeV,\quad m_{u,d} = 0 MeV, \quad m_s = 170 MeV;
 \]
The current masses are supposed to remain still in the
calculations when the temperature changes. Respectively, the
components are considered as ideal Fermi or Boson gases,
 \[
 f_i(p) = \frac{\omega_i}{e^{\frac{E_i-\mu_i}{T}}\pm 1},
 \]

Combining Equations
(\ref{eqs:bound})(\ref{eqs:yop})(\ref{eqs:yop0})(\ref{eqs:point})(\ref{eqs:bag_mass})(\ref{eqs:bagExt})(\ref{eq:de})
and strange quantum number conservation, one gets
 \begin{eqnarray}
 \nonumber m_i &=& m(\Delta\varepsilon),\\
 \nonumber V_i &=& V(\Delta\varepsilon),\\
 y_{op} &=& y_{op}(m_i, V_i, T, \mu_B, \mu_s),
 \label{eqs:selvesEQSA}\\
 \nonumber \Delta\varepsilon &=& \Delta\varepsilon(m_i, y_{op}, T, \mu_B, \mu_s),\\
 \nonumber \mu_s &=& \mu_s(m_i, y_{op}, T, \mu_B, \mu_s),
 \end{eqnarray}
where $\Delta\varepsilon = \Delta\varepsilon_{med}$ and $\mu_s$ is
the strange quark chemical potential,
 \[
 \mu_s=\mu_q-\mu_S=\frac{\mu_B}{3}-\mu_S,
 \]

 Eqs (\ref{eqs:selvesEQSA}) is a set of self-consistent
 equations equivalent to the forms with $\mu_B$ and $\Delta\varepsilon$ below,
 \begin{eqnarray*}
 \mu_s = \varphi_1(T, \Delta\varepsilon, \mu_B, \mu_s),\\
 \Delta\varepsilon = \varphi_2(T, \Delta\varepsilon, \mu_B, \mu_s),
 \end{eqnarray*}

 Setting $\mu_B=0$, one has,
 \begin{equation}
 \Delta\varepsilon = \varphi(T, \Delta\varepsilon),
 \label{eqs:self}
 \end{equation}
In the calculations, $gg$ states are neglected as their heavy
masses as "Glueballs"\cite{sign:sign GlueBall1}. Contributions of
$qg$ states are deduced by treating them as specific diquarks. For
simplicity, $E_M$ and $E_E$ are neglected from Eq
(\ref{eqs:bagExt}), when particle radii are larger than $16$
$GeV^{-1}$.

\begin{figure}[h]
 \centering
  \hspace{0.025\textwidth}%
  \begin{minipage}[t]{0.45\textwidth}
  \centering
  \includegraphics[width=\textwidth]{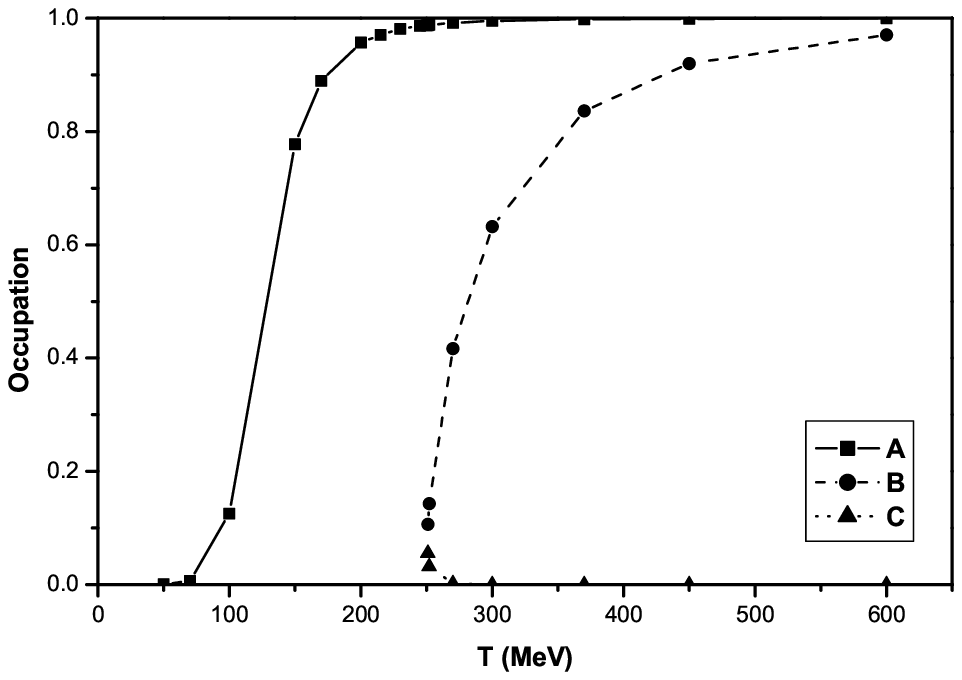}
  \caption{\quad Ratio of occupation at different Temperatures,
   where $C_\alpha =0.8$ and $C_e =5\times10^{-10}MeV^{-4}$.
   Data below $T_c$ is just to show the intention of the sequences.}
  \label{fig:OpRoot1}
  \end{minipage}%
  \hspace{0.05\textwidth}%
  \begin{minipage}[t]{0.45\textwidth}
  \centering
  \includegraphics[width=\textwidth]{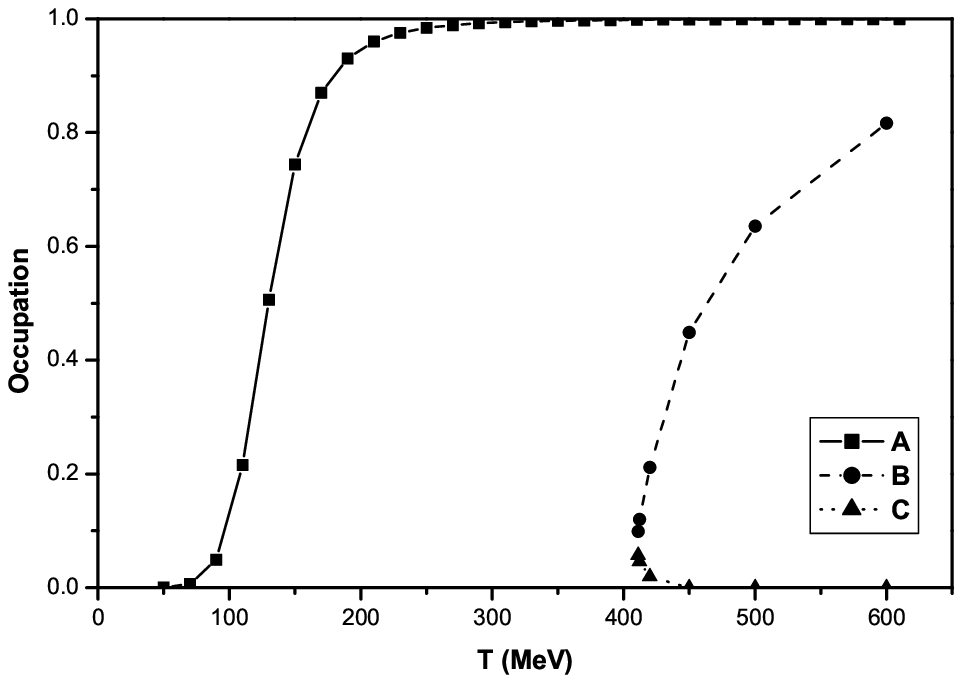}
  \caption{\quad Ratio of occupation at different Temperatures,
   where $C_\alpha =0.8$ and $C_e =1\times10^{-10}MeV^{-4}$.}
  \label{fig:OpRoot2}
  \end{minipage}%
  \hspace{0.025\textwidth}
\end{figure}

\begin{figure}[h]
 \centering
  \hspace{0.025\textwidth}%
  \begin{minipage}[t]{0.45\textwidth}
    \centering
    \includegraphics[width=\textwidth]{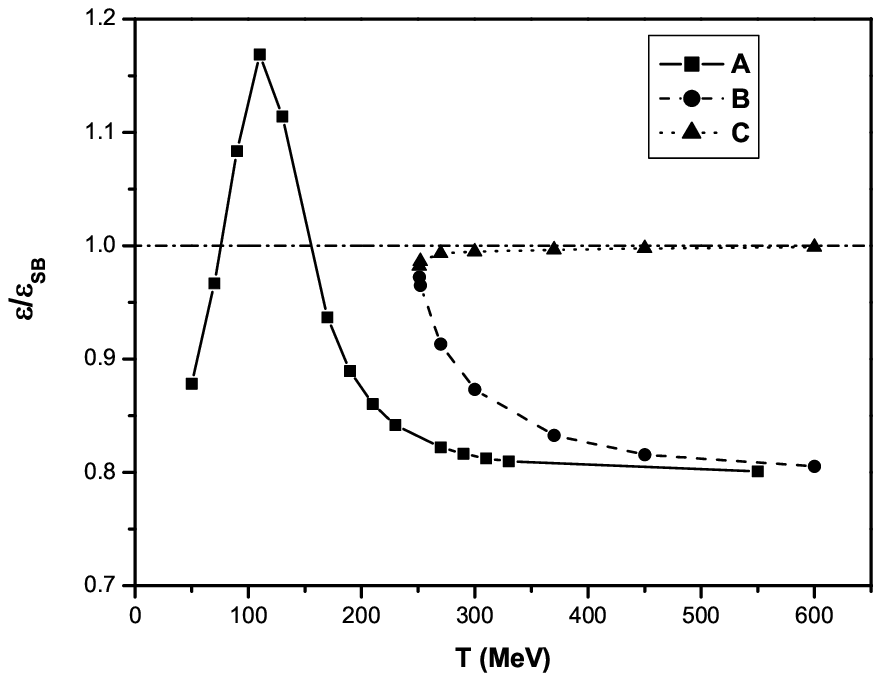}
    \caption{\quad Energy density of the QGP with binary bound states at different Temperatures,
     where $C_\alpha =0.8$ and $C_e =5\times10^{-10}MeV^{-4}$.}
    \label{fig:root_e1}
  \end{minipage}%
  \hspace{0.05\textwidth}%
  \begin{minipage}[t]{0.45\textwidth}
  \centering
    \includegraphics[width=\textwidth]{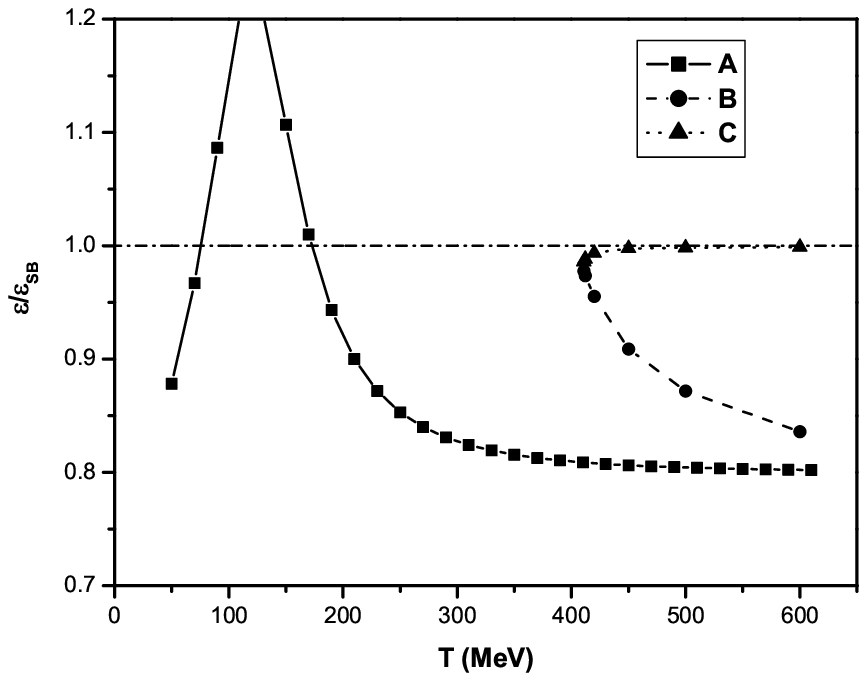}
    \caption{\quad Energy density of the QGP with binary bound states at different Temperatures,
     where $C_\alpha =0.8$ and $C_e =1\times10^{-10}MeV^{-4}$.}
    \label{fig:root_e2}
 \end{minipage}%
  \hspace{0.025\textwidth}
\end{figure}

\begin{figure}[htb]
 \centering
  \hspace{0.025\textwidth}%
  \begin{minipage}[t]{0.45\textwidth}
    \centering
    \includegraphics[width=\textwidth]{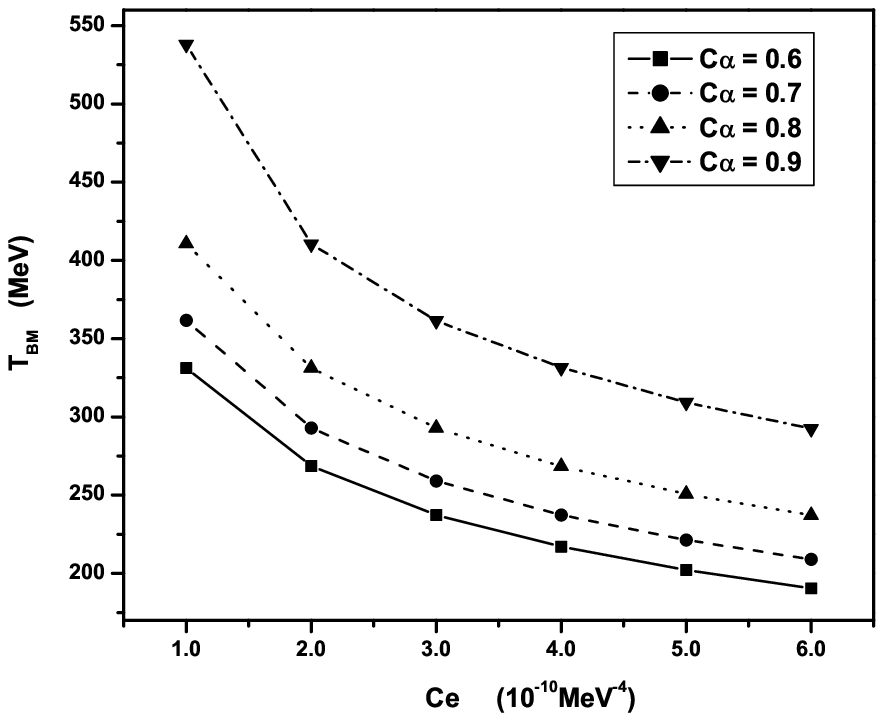}
    \caption{\quad $T_{BM}$ dependence on parameters of $C_\alpha$ and $C_e$.}
    \label{fig:root_para1}
  \end{minipage}%
  \hspace{0.05\textwidth}%
  \begin{minipage}[t]{0.45\textwidth}
  \centering
    \includegraphics[width=\textwidth]{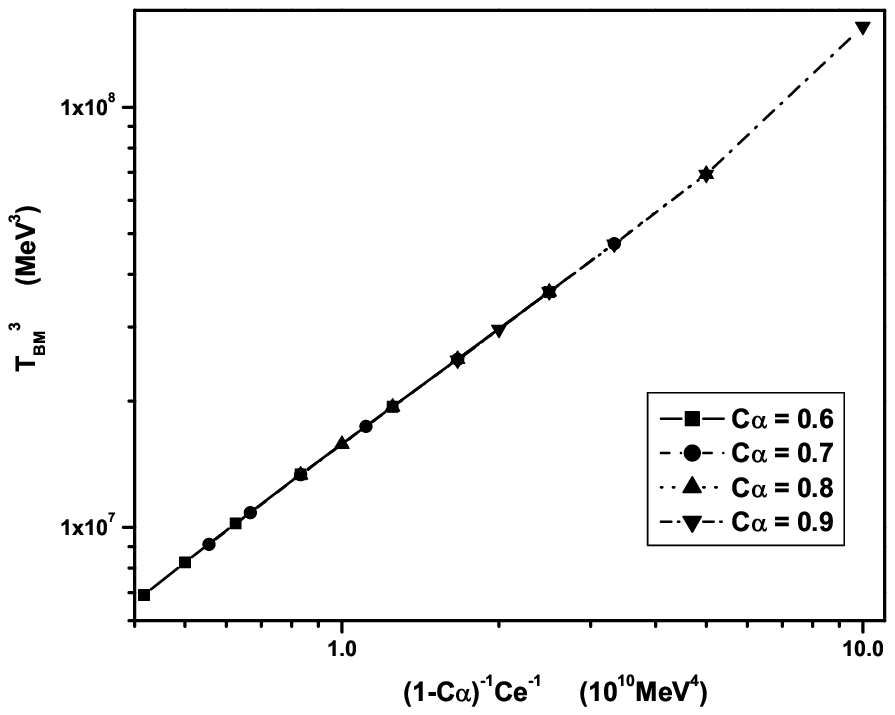}
    \caption{\quad $T_{BM}$ dependence on parameter of $(1-C_\alpha)C_e$.}
    \label{fig:root_paralinear}
 \end{minipage}%
  \hspace{0.025\textwidth}
\end{figure}

It is noted that there are only two adjustable parameters,
$C_\alpha$ and $C_e$, in our calculations, which makes the results
more comprehensive. A demonstration of the solutions of Eq
(\ref{eqs:self}) is shown in Fig (\ref{fig:RootFind}). The ratio
of occupations and energy density from the solutions under
different parameters are presented in Fig
(\ref{fig:OpRoot1})-(\ref{fig:root_e2}). As shown in Fig
(\ref{fig:RootFind}), there is only one root in the left at low
temperatures, which corresponds to the sequence "A" in Fig
(\ref{fig:OpRoot1})-(\ref{fig:root_e2}). When $T$ grows up beyond
a specific temperature $T_{BM}$, another two roots (corresponding
to sequences of "B" and "C") will appear. It is noted that the
sequence "B" is unstable. $T_{BM}$ increases with $C_\alpha$ and
decreases with $C_e$, as shown in Fig (\ref{fig:root_para1}). Data
here are normalized by the Stephen-Boltzmann Limit of ideal QGP,
which is defined as

 \begin{equation}
 \varepsilon_{SB} = (16+\frac{21}{2}n_f)\frac{\pi^2}{30}T^4;
 \end{equation}

It is easy to prove that the Free Energy of sequence "C" is
smaller than sequence "A". That means sequence "A" is a metastable
state when $T>T_{BM}$ like Superheat Liquids. There might be an
upper limit of temperature in the strongly coupled Quark Gluon
Plasma to melt out the bound states somewhere at $T>T_{BM}$.

According to Eq (\ref{eq:fe}) and (\ref{eq:de}), one could find
$(1-C_\alpha)C_e$ is major term in Eq (\ref{eq:fe}). That makes
$T_{BM}$ nearly a function of $(1-C_\alpha)C_e$ with a small
correction less than $1MeV$. It can be found clearly in Fig
(\ref{fig:root_para1}) and (\ref{fig:root_paralinear}). Detailed
investigation on the behavior of $T_{BM}$ implies that $T_{BM}^3$
is linear to $[(1-C_\alpha)C_e]^{-1}$ approximately in the region
of $0.4\times10^{10}MeV^4 < [(1-C_\alpha)C_e]^{-1} <
5\times10^{10}MeV^4$. Calculations with larger parameters than
$5\times10^{10}MeV^4$ may contain much larger uncertainties and
instabilities.

Details of the transition between sequences depend on the local
fluctuations, which are not clear so far. Approximately,
 \[
 \frac{\sigma_\varepsilon^2}{\varepsilon_{SB}^2} \sim O \left(\frac{4T\varepsilon}{V_L\varepsilon_{SB}^2}\right)
              \sim O
              \left(\frac{4\eta}{V_L\zeta_{SB}T^3}\right),
 \]
where $V_L$ is the concerning local volume, and
 \[
               \eta = \varepsilon/\varepsilon_{SB}, \qquad \zeta_{SB} =
              \varepsilon_{SB}/T^4
              =\frac{\pi^2}{30}\left(16+\frac{21}{2}N_f\right).
 \]

\begin{figure}[h]
  \centering
  \includegraphics[width=0.5\textwidth]{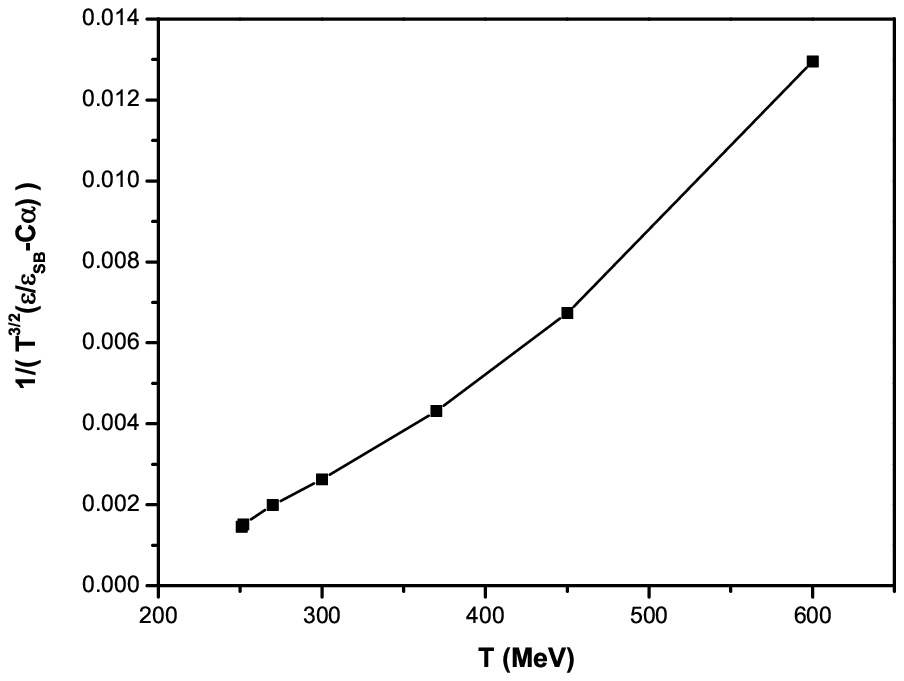}
  \caption{\quad Estimation of the local fluctuations of the metastable state (sequence "A").
  $C_\alpha =0.8$, $C_e =5\times10^{-10}MeV^{-4}$.}
  \label{fig:fluctuation}
\end{figure}

The influence of local fluctuations on sequence "A" could be
described as,
 \begin{equation}
 \frac{\sigma_\varepsilon}{\varepsilon_{B}-\varepsilon_{A}}
    \approx \frac{\sigma_\varepsilon}{\varepsilon_{B}-C_\alpha\varepsilon_{SB}}
    \propto \frac{1}{T^{3/2}(\varepsilon_{B}/\varepsilon_{SB}-C_\alpha)}
 \end{equation}
which increases with temperature ( shown in Fig
(\ref{fig:fluctuation}) ). It implies that when
$T\longrightarrow\infty$, the system should remain at sequence "C"
finally. If the local fluctuation is large enough at $T_{BM}$,
there might be a stage of mixed states from $T_{BM}$ before the
whole system transform to sequence "C". In this stage, local
states could evolve to each other between sequences "A" and "C" by
fluctuations. Otherwise, if the local fluctuation is too small to
drive the sequence "A" to leave the metastable region at $T_{BM}$,
a sudden conversion may occur somewhere at $T>T_{BM}$ when the
local fluctuation grows large enough.

As shown in Fig (\ref{fig:root_e1}) and (\ref{fig:root_e2}), the
energy density of the whole system may be larger than those of
heavy quasiparticles or conventional QGP in some regions. It is
even larger than the Stephen-Boltzmann Limit when $T$ is close to
$Tc$. Since the detailed behavior of color deconfinement is not
discussed here, the data close to $Tc$ are just qualitative. One
of the obvious raisons to enhance the energy density is that more
degrees of freedom are inserted into the system. Another raison is
that more light and invariable current quark masses are used in
the calculations instead of those ones proportional to
$T$\cite{sign:sign Latt_Pressure}. Accordingly, data below $Tc$
are not physical. They are only listed in the figures to show the
inclination of the sequences. When the temperature increases, the
ratio of occupations grows up, which could reduce the energy
density to a factor near $C_\alpha$ relative to
$\varepsilon_{SB}$, before the state transition takes place at a
limit temperature $T \geq T_{BM}$. After the conversion, bound
states will be melted out and the ratio of occupations will
decrease to zero. The energy density increases and then the system
restores to ideal QGP. Thinking that $C_\alpha$ may increase with
temperature, (because the reduction of phase space in the "ball"
is mainly subjected to low momentum regions,) the transition may
be more smooth than expected. If $C_\alpha$ grows to $1$
somewhere, bound states will exist for a long period and the
transition may vanish at all.

\vspace{10pt}

\leftline{\large \bf 5. Summary}

\vspace{8pt}

Considering the volume of bound states, we provided a
phenomenological mechanism to discuss its influence on the
properties of QGP. There are different solutions for the masses
and volumes of bound states in different conditions\cite{sign:sign
Latt_Radius}. We constructed an extended Bag Model as a
simplification to describe the volume effect qualitatively.
Although the details of the model may not be serious, it is still
helpful for us to understand the statistical behaviors of the
bound states with excluded volumes in QGP.

The contribution of the bound states to the system could be
described by the solutions of a self-consistent equation
(\ref{eqs:self}). From its numerical calculation, we found that
the energy density near $Tc$ is higher than the one with only
quarks and gluons, and the system is nearly occupied by the bound
states. The ratio of occupations increases with $T$, and the
energy density relative to $\varepsilon_{SB}$ or ideal gas is
reduced. When the temperature is larger than $T_{BM}$, multi-roots
of Eq(\ref{eqs:self}) start to appear. A conversion takes place in
the system to melt all the bound states at the limit temperature
$T \geq T_{BM}$.

\vspace{12pt}

\leftline{\large \bf Acknowledgement }

\vspace{8pt}

This work was supported in part by the National Natural Science
Foundation of China 90103019 and 10428510. We thank Professor
Yuxin Liu for useful discussions.


\vspace{8pt}

\begin{thebibliography}{99} \small

\bibitem[1]{sign:sign diquark_baryon} H. Miao, Z. B. Ma and C. S. Gao, J. Phys. {\bf{G29}} (2003) 2187-2192.

\bibitem[2]{sign:sign diquark_density} Z. B. Ma, H. Miao and C. S. Gao, Chin. Phys. Lett. {\bf{20}} (2003) 1691-1693.

\bibitem[3]{sign:sign bound_state}  E. V. Shuryak and I. Zahed,
Phys. Rev. {\bf D70} (2004) 054507, hep-ph/0403127.

\bibitem[4]{sign:sign QuasiParticle1} P. Petreczky, F. Karsch, E. Laermann, S.
Stickan and I. Wetzorke, Nucl. Phys. Proc. Suppl. {\bf 106} (2002)
513-515, hep-lat/0110111.

\bibitem[5]{sign:sign WaalsSPS} G. Zeeb, K. A. Bugaev, P. T. Reuter and H.
Stocker, nucl-th/0209011.

\bibitem[6]{sign:sign WaalsHG} M. I. Gorenstein, A. P. Kostyuk and Y. D.
Krivenko,  J. Phys. {\bf G25} (1999) L75-L83, nucl-th/9906068.

\bibitem[7]{sign:sign_vol_lit} J.~W.~Clark, J.~Cleymans and J.~Rafelski,
   Phys.\ Rev.\ {\bf C33}, 703 (1986).

\bibitem[8]{sign:sign EOS} P. Huovinen, Nucl. Phys. {\bf A761} (2005) 296-312, nucl-th/0505036.

\bibitem[9]{sign:sign byinQGP1} J. F. Liao and E. V. Shuryak, hep-ph/0508035.

\bibitem[10]{sign:sign Strange} C. S. Gao and T. Wu, J. Phys. {\bf G27} (2001) 459-463.

\bibitem[11]{sign:sign MIT_Bage} A. Chodos, et al., Phys. Rev. {\bf{D9}} (1974)
3471-3495.

\bibitem[12]{sign:sign Bag_mass} T. A. DeGrand, et al., Phys. Rev. {\bf{D12}} (1975) 2060-2076.

\bibitem[13]{sign:sign Bag_Liu} Y. X. Liu, D. F. Gao and H. Guo,
Nucl. Phys. {\bf{A695}} (2001) 353-364.

\bibitem[14]{sign:sign Bag_Burgio} G. F. Burgio, et al., Phys.
Lett. {\bf B526}, (2002) 19-26.

\bibitem[15]{sign:sign Bag_Aquirre} R. Aguirre, Phys. Lett. {\bf
B559}, (2003) 207-213.

\bibitem[16]{sign:sign hydro1} P. F. Kolb and U. W. Heinz, Invited
review for 'Quark Gluon Plasma 3', Editors: R. C. Hwa and X. N.
Wang, World Scientific, Singapore, nucl-th/0305084.

\bibitem[17]{sign:sign diquark_form} M. I. Pavkovi$\acute{c}$, Phys. Rev. {\bf D13} (1976) 2128-2138.

\bibitem[18]{sign:sign GlueBall1} J. C. Su, J. X. Chen, Phys. Rev. {\bf D69} (2004)
076002, hep-ph/0506113.

\bibitem[19]{sign:sign Latt_Pressure} F. Karsch, E. Laermann and A.
Peikert, Phys. Lett. {\bf B478} (2000) 447-455, hep-lat/0002003.

\bibitem[20]{sign:sign Latt_Radius} C. Alexandrou, P. de Forcrand and B. Lucini, Talk presented at Lattice 2005, hep-lat/0509113.

\end{thebibliography}
\end{document}